\documentclass[prd,showpacs,preprintnumbers,amsmath,amssymb]{revtex4}

\usepackage{graphics}
\usepackage{epsfig}
\usepackage{subfigure}
\usepackage{dcolumn}
\usepackage{bm}
\usepackage{color}

\begin{document}
\draft

\title{{\bf Possible Molecular States of $D^{*}_s\bar{D}^{*}_s$ System and $Y(4140)$ }}
\author{Gui-Jun Ding}

\affiliation{\centerline{Department of Modern
Physics,}\centerline{University of Science and Technology of
China,Hefei, Anhui 230026, China}}

\begin{abstract}

The interpretation of $Y(4140)$ as a $D^{*}_s\bar{D}^{*}_s$ molecule
is studied dynamically in the one boson exchange approach, where
$\sigma$, $\eta$ and $\phi$ exchange are included. Ten allowed
$D^{*}_s\bar{D}^{*}_s$ states with low spin parity are considered,
we find that the $J^{PC}=0^{++}$, $1^{+-}$, $0^{-+}$, $2^{++}$ and
$1^{--}$ $D^{*}_s\bar{D}^{*}_s$ configurations are most tightly
bound. We suggest the most favorable quantum numbers are
$J^{PC}=0^{++}$ for $Y(4140)$ as a $D^{*}_s\bar{D}^{*}_s$ molecule,
however, $J^{PC}=0^{-+}$ and $2^{++}$ can not be excluded. We
propose to search for the $1^{+-}$ and $1^{--}$ partners in the
$J/\psi\eta$ and $J/\psi\eta'$ final states, which is an important
test of the molecular hypothesis of $Y(4140)$ and the reasonability
of our model. The $0^{++}$ $B^{*}_s\bar{B}^{*}_s$ molecule is deeply
bound, experimental search in the $\Upsilon(1S)\phi$ channel at
Tevatron and LHC is suggested.

\end{abstract}

\maketitle

\vskip0.5cm

\section{Introduction}

Recently the CDF Collaboration reported a narrow structure $Y(4140)$
in the decay $B^{+}\rightarrow K^{+}Y(4140)$ followed by
$Y(4140)\rightarrow J/\psi\phi$ with a statistical significance of
3.8 $\sigma$. The mass and the width of this structure are fitted to
be $4143.0\pm2.9(\rm stat)\pm1.2(\rm syst)$ MeV and
$11.7^{+8.3}_{-5.0}(\rm stat)\pm3.7(\rm syst)$ MeV respectively
\cite{Aaltonen:2009tz,Yi:2009pz}. Since the quantum numbers of both
$J/\psi$ and $\phi$ are $I^{G}(J^{PC})=0^{-}(1^{--})$, $Y(4140)$ is
an isospin singlet with positive $C$ parity.

$Y(4140)$ is very similar to the charmonium like state $Y(3940)$,
which is observed in $B\rightarrow KY(3940)\rightarrow
KJ/\psi\omega$ \cite{belle-y3930,babar-y3930}. It was argued by the
CDF Collaboration that $Y(4140)$ can not be a conventional
charmonium state, because a $c\bar{c}$ charmonium state with mass
about 4143 MeV would dominantly decay into open charm pairs, and the
branch ratio into the double OZI forbidden modes $J/\psi\phi$ or
$J/\psi\omega$ is negligible. Although the decay $Y(4140)\rightarrow
J/\psi\phi$ is unusual, the decay
$\chi_{b1,2}(2P)\rightarrow\omega\Upsilon(1S)$ has been observed
\cite{Severini:2003qw}. Therefore $Y(4140)$ as a conventional
$c\bar{c}$ state can not be ruled out completely due to the scarcity
of current experiment data. Comparing with the theoretical
predictions for the charmonium spectrum
\cite{Barnes:2005pb,Godfrey:1985xj}, we suggest that $Y(4140)$ would
most likely to be the $2\,^1D_2$ state with
$I^{G}(J^{PC})=0^{+}(2^{-+})$, if it is a $c\bar{c}$ charmonium
state. It would dominantly decay into $DD^{*}$, $D^{*}D^{*}$ and
$D_sD^{*}_s$ in this scenario. A possible explanation for the
unusual decay mode $Y(4140)\rightarrow J/\psi\phi$ is that the
rescattering through $D_s\bar{D}^{*}_s/D^{*}_s\bar{D}_s$ or
$D^{*}_s\bar{D}^{*}_s$ may be responsible, i.e. $Y(4140)\rightarrow
D_s\bar{D}^{*}_s(D^{*}_s\bar{D}_s\;{\rm
or}\;D^{*}_s\bar{D}^{*}_s)\rightarrow J/\psi\phi$. Mixing between
the charmonium state and the $D^{*}_s\bar{D}^{*}_s$ molecule may
also contribute to this extraordinary decay.

There are already some theoretical discussions about this structure.
The authors in Ref. \cite{Liu:2009ei} suggested that $Y(4140)$ is a
molecular partner of $Y(3940)$, and that it is a $J^{PC}=0^{++}$ or
$2^{++}$ $D^{*}_s\bar{D}^{*}_s$ molecule. In Ref.
\cite{Mahajan:2009pj}, the author argued that $Y(4140)$ is a
$D^{*}_s\bar{D}^{*}_s$ molecule or a exotic $1^{-+}$ charmonium
hybrid. The interpretation of $Y(4140)$ as a $0^{++}$
$D^{*}_s\bar{D}^{*}_s$ molecule has been studied by QCD sum rules
\cite{Albuquerque:2009ak,Wang:2009ue}, however, different
conclusions were reached. The decay of $Y(4140)$ as a hadronic
molecule or conventional charmonium $\chi'_{c0,1}$ were discussed as
well \cite{Branz:2009yt,Liu:2009iw}.

In the past years, some new states near threshold have been
observed, and they are usually suggested to be hadronic molecules
\cite{
Close:2003sg,Voloshin:2003nt,Wong:2003xk,swanson,Tornqvist:2004qy,Ding:2008mp,Ding:2008gr}.
Being different from familiar hadronic molecule candidates,
$Y(4140)$ is about 81.6 MeV below the $D^{*}_s\bar{D}^{*}_s$
threshold, its binding energy is not small if it is identified as a
$D^{*}_s\bar{D}^{*}_s$ molecule. There are various theoretical
methods to study the hadronic molecule dynamically, such as the QCD
sum rule, the QCD effective field theory \cite{zhu_liu,Ding:2008gr},
the potential models with pairwise interactions between quarks
\cite{Wong:2003xk,Ding:2008mp} and so on. Tornqvist's original work
on hadronic molecule from one pion exchange is especially attractive
\cite{Tornqvist:1991ks,Tornqvist:1993ng}. Guided by the binding of
deuteron, he performed a systematic study of possible deuteronlike
two mesons bound states with long distance one pion exchange. Being
different from the above mentioned theoretical methods, he took into
account the contribution of the tensor force. Since the tensor force
turns out to be very important in the deuteron binding, one expects
that it would make a significant contribution to the binding of the
general hadronic molecule. At short distance, the interaction should
be induced by the interactions among the quarks in principle.
However, a detailed and reliable modeling of the short range
interaction is not an easy matter, and various phenomenological
models have been proposed \cite{Barnes:1991em,Swanson:1992ec},
although one pion exchange is expected to be dominant for the
hadronic molecule. Inspired by the nucleon-nucleon interactions, we
further extended the one pion exchange model to represent the short
distance contributions by the heavier mesons $\eta$, $\sigma$,
$\rho$ and $\omega$ exchanges in Ref. \cite{Ding:2009vj}. From these
studies, we learn that one pion exchange really dominates the
physics of hadronic molecule, and the tensor forces indeed plays a
critical role in providing the binding, which results in the S-D
wave mixing. Since in general the molecular state is weakly bound,
the separation between the two hadrons in the molecule should be
large. Consequently the dominance of one pion exchange can be easily
understood. A remarkable success of the model is its prediction for
the existence of the famous X(3872) long ago
\cite{Tornqvist:2004qy,Tornqvist:1993ng}. Based on one pion
exchange, Frank Close et al. recently suggested that the charmed
mesons molecular state is a possible solution to the enigmatic
states observed above 4 GeV in $e^{+}e^{-}$ annihilation
\cite{Close:2009ag}. Therefore we think that the one boson exchange
model can give a reasonable and reliable description for the
hadronic molecule. In this work, we shall study the possible
molecular states of the $D^{*}_s\bar{D}^{*}_s$ system and discuss
the most favorable quantum numbers of $Y(4140)$ as a
$D^{*}_s\bar{D}^{*}_s$ molecule in the one boson exchange model. Ten
allowed $D^{*}_s\bar{D}^{*}_s$ states with total angular momentum
smaller than 3 will be discussed.

The paper is organized as follows. In section II, the effective
potentials of the ten $D^{*}_s\bar{D}^{*}_s$ molecular states are
given. The numerical results and the possible $D^{*}_s\bar{D}^{*}_s$
molecules are presented in section III. Section IV discusses the
most favorable quantum numbers of $Y(4140)$ and the implications, if
it is interpreted as a $D^{*}_s\bar{D}^{*}_s$ molecule. Finally we
give our conclusions and some discussions in section V.

\section{The effective potentials for the $D^{*}_s\bar{D}^{*}_s$ states with low spin parity}

In the one boson exchange model
\cite{Tornqvist:1991ks,Tornqvist:1993ng,Ding:2009vj} , the effective
potentials between two hadrons are obtained by summing the
interactions between light quarks or antiquarks via one boson
exchange, and one boson exchange only between up quarks or down
quarks has been considered. We note that the present one boson
exchange is different from the well-known Goldstone boson exchange
constituent quark model \cite{Glozman:1997ag}, where one boson
exchange between quarks inside a hadron is used to describe the
light baryon spectrum. Since strange quark is involved in the
present problem, we should extend our consideration to the one boson
exchange between strange quarks. Because $D^{*}_s$ is an isospin
singlet, the effective potential between $D^{*}_s$ and
$\bar{D}^{*}_s$ is induced by $\eta$, $\sigma$ and $\phi$ exchanges,
whereas the $\pi$, $\rho$ and $\omega$ exchanges don't contribute.
Taking into account the overall sign difference $(-1)^{G}$ between
the quark-quark interactions and the quark-antiquark interactions,
where $G$ is the $G$-parity of the exchanged meson, we obtain the
effective potential
\begin{equation}
\label{1}V(r)=V_{\eta}(r)+V_{\sigma}(r)-V_{\phi}(r)\equiv
V_{C}(r)+V_{S}(r)\,\bm{\sigma}_i\cdot\bm{\sigma}_j+V_{T}(r)S_{ij}(\hat{\mathbf{r}})+V_{LS}(r)\mathbf{L}\cdot\mathbf{S}_{ij}
\end{equation}

\begin{figure}
\includegraphics[scale=.745]{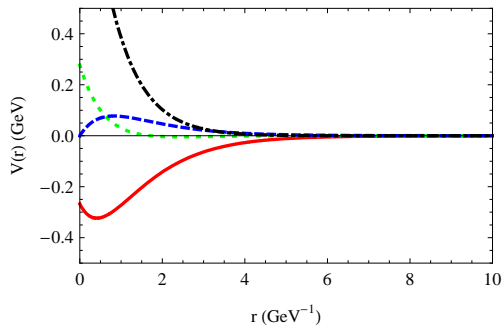}
\caption{\label{potential} The interaction potentials $V_C(r)$,
$V_{S}(r)$, $V_T(r)$ and $V_{LS}(r)$ with $\Lambda=1400$ MeV. The
solid line represents the central potential $V_C(r)$, the dotted,
dashed and dash-dotted lines denote $V_{S}(r)$, $V_T(r)$ and
$V_{LS}(r)$ respectively.}
\end{figure}

where $V_{\eta}(r)$, $V_{\sigma}(r)$ and $V_{\phi}(r)$ respectively
denote the effective potentials from $\eta$, $\sigma$ and $\phi$
exchange between two quarks, and their expressions are given
explicitly in Eq.(9)-Eq.(11) of Ref. \cite{Ding:2009vj}. Noting that
the $\phi$ exchange distinguishes the $D^{*}_s\bar{D}^{*}_s$ from
the exotic $D^{*}_s{D}^{*}_s$ systems. The subscript $i$ and $j$ are
the indexes of the strange quark and strange antiquark.
$S_{ij}(\hat{\mathbf{r}})=3(\bm{\sigma}_i\cdot\hat{\mathbf{r}})(\bm{\sigma}_j\cdot\hat{\mathbf{r}})-\bm{\sigma}_i\cdot\bm{\sigma}_j$
is the well-known tensor operator,
$\mathbf{S}_{ij}=\frac{1}{2}(\bm{\sigma}_i+\bm{\sigma}_j)$ is the
total spin of light quarks, and $\mathbf{L}$ is the relative angular
momentum operator. Comparing with the general results in Ref.
\cite{Ding:2009vj}, the isospin dependent interactions vanish here.
We would like to note that the spin operator $\bm{\sigma}_i$ or
$\bm{\sigma}_j$ only acts on the strange quark or antiquark here.
The potentials $V_C(r)$, $V_S(r)$, $V_{T}(r)$ and $V_{LS}(r)$ are
given as follows:
\begin{eqnarray}
\nonumber&& V_C(r)=-\frac{g^2_{\sigma ss}}{4\pi}\,m_{\sigma}H_0(\Lambda,m_{\sigma},r)-\frac{g^2_{\sigma ss}}{4\pi}\frac{m^3_{\sigma}}{8m^2_s}H_1(\Lambda,m_{\sigma},r)-\frac{g^2_{\phi ss}}{4\pi}\,m_{\phi}H_0(\Lambda,m_{\phi},r)+
\frac{g^2_{\phi ss}+4g_{\phi ss}f_{\phi ss}}{4\pi}\frac{m^3_{\phi}}{8m^2_s}H_1(\Lambda,m_{\phi},r)\\
\nonumber&&V_S(r)=-\frac{g^2_{\eta ss}}{4\pi}\frac{m^3_{\eta}}{12m^2_s}H_1(\Lambda,m_{\eta},r)+\frac{(g_{\phi ss}+f_{\phi ss})^2}{4\pi}\frac{m^3_{\phi}}{6m^2_s}H_1(\Lambda,m_{\phi},r)\\
\nonumber&&V_{T}(r)=\frac{g^2_{\eta ss}}{4\pi}\frac{m^3_{\eta}}{12m^2_s}H_3(\Lambda,m_{\eta},r)+\frac{(g_{\phi ss}+f_{\phi ss})^2}{4\pi}\frac{m^3_{\phi}}{12m^2_s}H_3(\Lambda,m_{\phi},r)\\
\label{2}&&V_{LS}(r)=-\frac{g^2_{\sigma ss}}{4\pi}\frac{m^3_{\sigma}}{2m^2_s}H_2(\Lambda,m_{\sigma},r)+\frac{3g^2_{\phi ss}+4g_{\phi ss}f_{\phi ss}}{4\pi}\frac{m^3_{\phi}}{2m^2_s}H_2(\Lambda,m_{\phi},r)
\end{eqnarray}
where $m_s$ is the mass of the constituent strange quark, and we
take $m_s=0.55$ GeV in this work. $m_{\eta}$, $m_{\sigma}$ and
$m_{\phi}$ are the masses of $\eta$, $\sigma$ and $\phi$ mesons
respectively. $g_{{\cal M}ss}$(${\cal M}=\eta$, $\sigma$ and $\phi$)
and $f_{\phi ss}$ denote the effective coupling constants between
the exchanged meson and the strange quarks. In the above formulae,
we have introduced form factor at each interaction vertex, and the
form factor in momentum space is taken as
\begin{equation}
\label{add1}F(q)=\frac{\Lambda^2-m^2}{\Lambda^2-q^2}
\end{equation}
where $\Lambda$ is the so-called regularization parameter, $m$ and
$q$ are the mass and the four momentum of the exchanged boson
respectively. The form factor suppresses the contribution of high
momentum, i.e. small distance. The presence of such a form factor is
dictated by the extended structure of the hadrons. The parameter
$\Lambda$, which governs the range of suppression, can be directly
related to the hadron size which is approximately proportional to
$1/\Lambda$. However, since the question of hadron size is still
very much open, the value of $\Lambda$ is poorly known
phenomenologically, and it is dependent on the models and
applications. In the nucleon-nucleon interactions, the $\Lambda$ in
the range of 0.8-1.5 GeV has been used to fit the data. The extended
structure of hadrons also has the following obvious consequence:
because the mass of the exchanged meson determines the range of the
corresponding contribution to the $D^{*}_s\bar{D}^{*}_s$
interactions, one should restrict oneself to meson exchange with the
exchanged meson mass below a certain value, typically on the order
of the regularization parameter $\Lambda$. The $\eta$, $\sigma$ and
$\phi$ exchange are considered in the present work, therefore the
value of $\Lambda$ should be at least larger than the $\phi$ meson
mass. The functions $H_0(\Lambda,m,r)$, $H_1(\Lambda,m,r)$,
$H_2(\Lambda,m,r)$ and $H_3(\Lambda,m,r)$ involved in Eq.(\ref{2})
are given by
\begin{eqnarray}
\nonumber H_0(\Lambda,m,r)&=&\frac{1}{m r}\big(e^{-m
r}-e^{-\Lambda r}\big)-\frac{\Lambda^2-m^2}{2m\Lambda}\,e^{-\Lambda r}\\
\nonumber H_1(\Lambda,m,r)&=&-\frac{1}{m r}\big(e^{-m
r}-e^{-\Lambda r}\big)+\frac{\Lambda(\Lambda^2-m^2)}{2m^3}\,e^{-\Lambda r}\\
\nonumber H_2(\Lambda,m,r)&=&\big(1+\frac{1}{m
r}\big)\frac{1}{m^2r^2}e^{-m r}-\big(1+\frac{1}{\Lambda
r}\big)\frac{\Lambda}{m}\frac{1}{m^2r^2}e^{-\Lambda
r}-\frac{\Lambda^2-m^2}{2m^2}\frac{e^{-\Lambda r}}{m
r}\\
\label{3}H_3(\Lambda,m,r)&=&\big(1+\frac{3}{m
r}+\frac{3}{m^2r^2}\big)\frac{1}{m r}e^{-m
r}-\big(1+\frac{3}{\Lambda
r}+\frac{3}{\Lambda^2r^2}\big)\frac{\Lambda^2}{m^2}\frac{e^{-\Lambda
r}}{m r}-\frac{\Lambda^2-m^2}{2m^2}\big(1+\Lambda
r\big)\frac{e^{-\Lambda r}}{m r}
\end{eqnarray}

\begin{center}
\begin{table}[hptb]
\begin{tabular}{|cl|}\hline\hline

$J^{PC}$&~~~~~States\\\hline

$0^{++}$    & ~~~~~$^1S_0$, $^5D_0$    \\
$1^{+-}$    & ~~~~~$^3S_1$,$^3D_1$   \\
$0^{-+}$    &~~~~~ $^3P_0$   \\
$1^{++}$    &~~~~~ $^5D_1$     \\
$1^{-+}$    &~~~~~ $^3P_1$    \\
$2^{+-}$    &~~~~~ $^3D_2$    \\
$1^{--}$    &~~~~~ $^1P_1$, $^5P_1$, $^5F_1$      \\
$2^{++}$    &~~~~~ $^1D_2$, $^5S_2$, $^5D_2$, $^5G_2$        \\
$2^{-+}$    & ~~~~~$^3P_2$, $^3F_2$      \\
$2^{--}$    &~~~~~ $^5P_2$, $^5F_2$      \\
\hline \hline
\end{tabular}
\caption{\label{states}The allowed states of the
$D^{*}_s\bar{D}^{*}_s$ system with low spin parity, where we cut off
the total angular momentum to $J=2$.}
\end{table}
\end{center}

The four components of the interaction potentials $V_C(r)$,
$V_{S}(r)$, $V_T(r)$ and $V_{LS}(r)$ are displayed in Fig.
\ref{potential}. We see that the central potential $V_C(r)$ is
negative, whereas the remaining three components are positive. These
potentials are the cancellation results of different contributions,
especially for the tensor potential $V_{T}(r)$, whose amplitude is
smaller than that of the other three potentials. For the
$D^{*}_s\bar{D}^{*}_s$ system, the spatial parity is determined by
$P=(-1)^{L}$ and the $C$-parity is $C=(-1)^{L+S}$, where $L$ is the
relative angular momentum between $D^{*}_s$ and $\bar{D}^{*}_s$, and
$S$ is the total spin of the system. We cut off the total angular
momentum to $J=2$, the allowed states with low spin parity are
listed in Table \ref{states}. The matrix elements of operators
$\bm{\sigma}_i\cdot\bm{\sigma}_j$, $S_{ij}(\hat{\mathbf{r}})$ and
$\mathbf{L}\cdot\mathbf{S}_{ij}$ can be calculated straightforwardly
with the help of angular momentum algebra, and the results are given
analytically in the appendix of Ref.\cite{Ding:2009vj}. Consequently
we can write out the one boson exchange potentials for the allowed
states in matrix form as follows
\begin{eqnarray}
\label{4}V_{0^{++}}(r)&=&\left(\begin{array}{cc}
1&0\\
0&1
\end{array}\right)V_C(r)+\left(\begin{array}{cc}
-2&0\\
0&1
\end{array}\right)V_S(r)+\left(\begin{array}{cc}
0&-\sqrt{2}\\
-\sqrt{2}&-2
\end{array}\right)V_T(r)+\left(\begin{array}{cc}
0&0\\
0&-3
\end{array}\right)V_{LS}(r)
\end{eqnarray}

\begin{eqnarray}
\label{5}V_{1^{+-}}(r)&=&\left(\begin{array}{cc}
1&0\\
0&1
\end{array}\right)V_C(r)+\left(\begin{array}{cc}
-1&0\\
0&-1
\end{array}\right)V_S(r)+\left(\begin{array}{cc}
0&\sqrt{2}\\
\sqrt{2}&-1
\end{array}\right)V_T(r)+\left(\begin{array}{cc}
0&0\\
0&-\frac{3}{2}
\end{array}\right)V_{LS}(r)
\end{eqnarray}

\begin{equation}
\label{6}V_{0^{-+}}(r)=V_C(r)-V_{S}(r)-2V_T(r)-V_{LS}(r)
\end{equation}

\begin{equation}
\label{7}V_{1^{++}}(r)=V_C(r)+V_S(r)-V_T(r)-\frac{5}{2}V_{LS}(r)
\end{equation}

\begin{equation}
\label{8}V_{1^{-+}}(r)=V_C(r)-V_S(r)+V_T(r)-\frac{1}{2}V_{LS}(r)
\end{equation}

\begin{equation}
\label{9}V_{2^{+-}}(r)=V_C(r)-V_S(r)+V_T(r)-\frac{1}{2}V_{LS}(r)
\end{equation}

\begin{eqnarray}
\label{10}V_{1^{--}}(r)=\left(\begin{array}{ccc}
1&0&0\\
0&1&0\\
0&0&1
\end{array}\right)V_C(r)+\left(\begin{array}{ccc}
-2&0&0\\
0&1&0\\
0&0&1
\end{array}\right)V_S(r)+\left(\begin{array}{ccc}
0&\frac{2}{\sqrt{5}}&-\sqrt{\frac{6}{5}}\\
\frac{2}{\sqrt{5}}&-\frac{7}{5}&\frac{\sqrt{6}}{5}\\
-\sqrt{\frac{6}{5}}&\frac{\sqrt{6}}{5}&-\frac{8}{5}
\end{array}\right)V_T(r)+\left(\begin{array}{ccc}
0&0&0\\
0&-\frac{3}{2}&0\\
0&0&-4
\end{array}\right)V_{LS}(r)
\end{eqnarray}

\begin{eqnarray}
\nonumber V_{2^{++}}(r)&=&\left(\begin{array}{cccc}
1&0&0&0\\
0&1&0&0\\
0&0&1&0\\
0&0&0&1
\end{array}\right)V_C(r)+\left(\begin{array}{cccc}
-2&0&0&0\\
0&1&0&0\\
0&0&1&0\\
0&0&0&1
\end{array}\right)V_S(r)+\left(\begin{array}{cccc}
0&-\sqrt{\frac{2}{5}}&\frac{2}{\sqrt{7}}&-\frac{6}{\sqrt{35}}\\
-\sqrt{\frac{2}{5}}&0&\sqrt{\frac{14}{5}}&0\\
\frac{2}{\sqrt{7}}&\sqrt{\frac{14}{5}}&\frac{3}{7}&\frac{12}{7\sqrt{5}}\\
-\frac{6}{\sqrt{35}}&0&\frac{12}{7\sqrt{5}}&-\frac{10}{7}
\end{array}\right)V_{T}(r)\\
\label{11}&&+\left(\begin{array}{cccc}
0&0&0&0\\
0&0&0&0\\
0&0&-\frac{3}{2}&0\\
0&0&0&-5
\end{array}\right)V_{LS}(r)
\end{eqnarray}

\begin{equation}
\label{12}V_{2^{-+}}(r)=\left(\begin{array}{cc}
1&0\\
0&1
\end{array}\right)V_C(r)+\left(\begin{array}{cc}
-1&0\\
0&-1
\end{array}\right)V_S(r)+\left(\begin{array}{cc}
-\frac{1}{5}&\frac{3\sqrt{6}}{5}\\
\frac{3\sqrt{6}}{5}&-\frac{4}{5}
\end{array}\right)V_T(r)+\left(\begin{array}{cc}
\frac{1}{2}&0\\
0&-2
\end{array}\right)V_{LS}(r)
\end{equation}

\begin{equation}
\label{13}V_{2^{--}}(r)=\left(\begin{array}{cc}
1&0\\
0&1
\end{array}\right)V_C(r)+\left(\begin{array}{cc}
1&0\\
0&1
\end{array}\right)V_S(r)+\left(\begin{array}{cc}
\frac{7}{5}&\frac{6}{5}\\
\frac{6}{5}&-\frac{2}{5}
\end{array}\right)V_T(r)+\left(\begin{array}{cc}
-\frac{1}{2}&0\\
0&-3
\end{array}\right)V_{LS}(r)
\end{equation}

In the following, we will perform the same analysis as for the
deuteron and the possible heavy flavor molecules in Ref.
\cite{Ding:2009vj}. One can then determine for which quantum numbers
the one boson exchange potential is attractive and strong enough so
that $D^{*}_s\bar{D}^{*}_s$ bound states are expected.

\section{Numerical results and possible $D^{*}_s\bar{D}^{*}_s$ molecules}

The input parameters in our model are the involved meson masses, the
effective coupling constants $g_{{\cal M}ss}$(${\cal M}=\eta$,
$\sigma$ and $\phi$) and $f_{\phi ss}$, and the regularization
parameter $\Lambda$. The meson masses are taken from the compilation
of the Particle Data Group \cite{pdg}: $m_{\eta}=547.853$ MeV,
$m_{\sigma}=600$ MeV, $m_{\phi}=1019.455$ MeV and
$m_{D^{*}_s}=2112.3$ MeV.  In the SU(3) flavor symmetry limit, the
coupling constants $g_{{\cal M}ss}$(${\cal M}=\eta$, $\sigma$ and
$\phi$) and $f_{\phi ss}$ between the exchanged mesons and strange
quarks are related to coupling constants $g_{{\cal M}qq}$(${\cal
M}=\eta$, $\sigma$ and $\phi$) and $f_{\phi qq}$ between the
exchanged mesons and up/down quarks via the following relations
\begin{eqnarray}
\label{14}g_{\eta ss}=-2g_{\eta qq},~~g_{\sigma ss}=g_{\sigma qq},~~g_{\phi ss}=\sqrt{2}\,g_{\omega qq},~~f_{\phi ss}=\sqrt{2}\,f_{\omega qq}
\end{eqnarray}
The coupling constants $g_{{\cal M}qq}$(${\cal M}=\eta$, $\sigma$
and $\omega$) and $f_{\omega qq}$ can be estimated from the
phenomenologically known $\eta NN$, $\sigma NN$ and $\omega NN$
coupling constants. Riska and Brown have explicitly demonstrated
that the nucleon resonance transition couplings to $\pi$, $\rho$ and
$\omega$ can be derived in the single quark operator approximation
\cite{Riska:2000gd}, which are in good agreement with the
experimental data. Adopting the same method, we can
straightforwardly derive the following relations between the
meson-quark couplings and the meson-nucleon couplings
\cite{Ding:2009vj,Riska:2000gd},
\begin{eqnarray}
\label{15} g_{\eta qq}=\frac{m_q}{m_{N}}g_{\eta NN},~~~g_{\sigma qq}=\frac{1}{3}\,g_{\sigma NN},~~~g_{\omega qq}=\frac{1}{3}\,g_{\omega
NN},~~~f_{\omega qq}=\frac{m_q}{m_N}f_{\omega
NN}-(\frac{1}{3}-\frac{m_q}{m_N})g_{\omega NN}
\end{eqnarray}
where $m_{N}$ is the nucleon mass. In this work, the effective
couplings between the exchanged bosons and the nucleons are taken
from from the well-known Bonn model \cite{Machleidt:1987hj}:
${g^{2}_{\eta NN}}/(4\pi)=3.0$, $g^2_{\sigma NN}/(4\pi)=7.78$,
$g^2_{\omega NN}/(4\pi)=20.0$ and $f^2_{\omega NN}/(4\pi)=0$. In
nature and in QCD, the flavor SU(3) symmetry is broken by non-equal
masses of the up and down quarks and the strange quark or the
electromagnetic effects. It is commonly believed that the error of
SU(3) predictions is approximately $20\%-30\%$. Consequently the
uncertainty of the coupling constants $g_{\eta ss}$, $g_{\sigma
ss}$, $g_{\phi ss}$ and $f_{\phi ss}$ is about $20\%-30\%$ as well.
As a demonstration for the consequence induced by the uncertainties
of the effective couplings, all the coupling constants would be
reduced by 20 percents later, and the corresponding predictions are
analyzed seriously. Taking into account the centrifugal barrier and
solving the coupled channel Schr$\ddot{\rm o}$dinger equations
numerically, then we can obtain the predictions for the binding
energy and the static properties, which are listed in the tables of
the Appendix. We notice that these predictions are rather sensitive
to the regularization parameter $\Lambda$ and the effective
couplings, this is common to the one boson exchange model
\cite{Liu:2009ei,Ding:2009vj,Thomas:2008ja}. We also find that the
binding energy increases with $\Lambda$, the reason is that
increasing $\Lambda$ increases the strength of the potential at
short distance.

For the $0^{++}$ $D^{*}_s\bar{D}^{*}_s$ state, the system can be in
$^1S_0$ or $^5D_0$ configuration, this is very similar to the
deuteron, which can be in $^3S_1$ or $^3D_1$. The S wave state mixes
with the D wave state under the tensor force, as is shown explicitly
in Eq.(\ref{4}). The energy of the system would be lowed
substantially due to the freedom of flipping from the $^1S_0$
configuration to the $^5D_0$ configuration and back. For $\Lambda$
in the range of 1350 MeV-1600 MeV, we can find a bound state with
the binding energy $\varepsilon=6.46-168.73$ MeV. The D wave
probability increases with the binding energy $\varepsilon$, and it
is about 12.73$\%$ for $\varepsilon=97.73$ MeV, the importance of
the tensor force is obvious. If all the coupling constants are
reduced by $20\%$, we need increase $\Lambda$ by about 200 MeV in
order to obtain similar binding energy. However, the value of
$\Lambda$ is still in the reasonable range. Since the molecular
state is widely extended, the decay into light mesons via
annihilation is generally suppressed by the form factor. The leading
source of decay is dissociation, to a good approximation the
dissociation will proceed via the free space decay of the
constituent mesons. Consequently the $0^{++}$ $D^{*}_s\bar{D}^{*}_s$
molecule mainly decays into $D^{+}_sD^{-}_s\gamma\gamma$, and
$D^{+}_sD^{-}_s\gamma\pi^{0}$, and the mode
$D^{+}_sD^{-}_s\pi^{0}\pi^{0}$ is forbidden by the phase space.

For the axial vector $1^{+-}$ state, there are two channels $^3S_1$
and $^3D_1$. The coupling between the S wave and D wave has the same
strength as the $0^{++}$ state, while the S wave spin-spin
interaction potential $V_S(r)$ is weaker than the corresponding one
of the $0^{++}$ state. Therefore the predictions for the binding
energy and the static properties have similar pattern with the
$0^{++}$ sector, and the binding energy of the $1^{+-}$ state is
somewhat smaller than that of the latter for the same $\Lambda$
value. We note that the unnatural spin parity forbids its decay into
$D_s\bar{D}_s$, while the decay mode
$D_s\bar{D}^{*}_s/D^{*}_s\bar{D}_s$ is allowed.

\begin{figure}[hptb]
\begin{center}
\begin{tabular}{ccc}
\includegraphics[scale=.645]{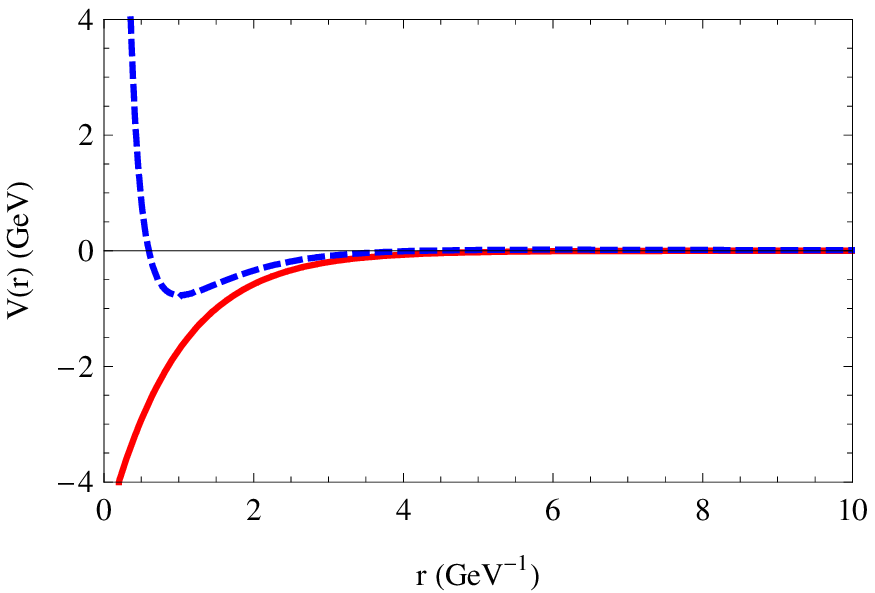}&\includegraphics[scale=.645]{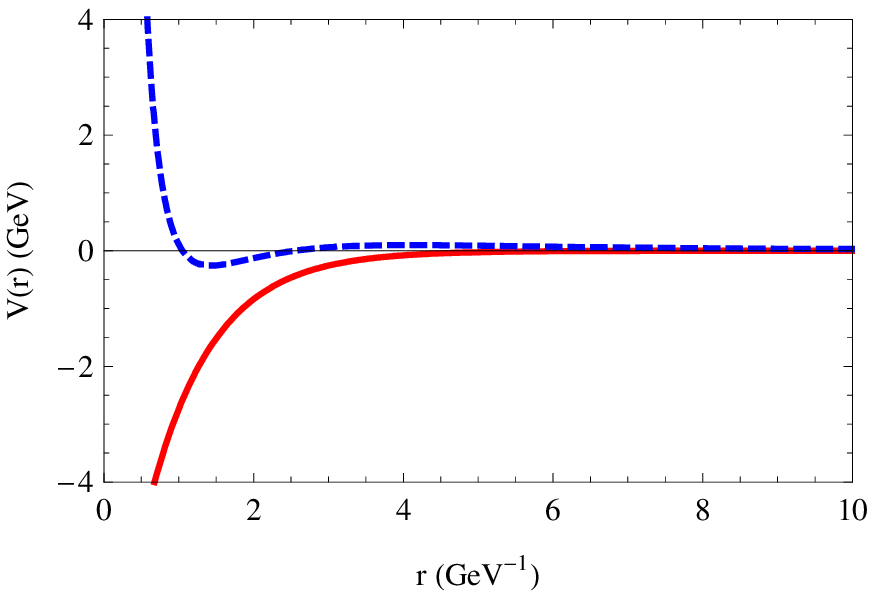}&\includegraphics[scale=.645]{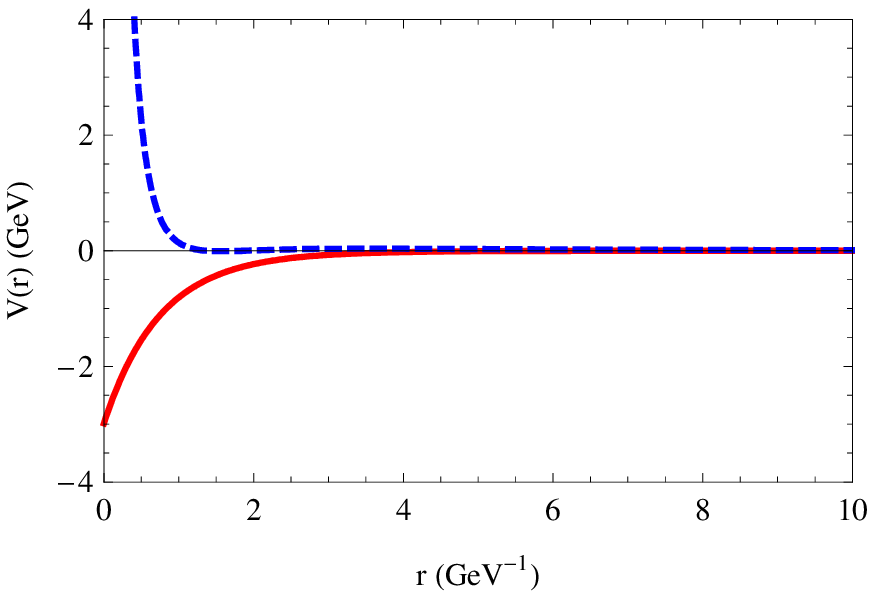}\\
(a)&(b)&(c)
\end{tabular}
\caption{\label{single_channel_potential}The potentials for the
single channel states with $\Lambda=1600$ MeV. (a), (b) and (c)
correspond to $0^{-+}$, $1^{++}$ and $1^{-+}$ states respectively.
The solid line represents the potential from one boson exchange, and
the dashed line denotes the effective potential after taking into
account the centrifugal barrier.}
\end{center}
\end{figure}

The $0^{-+}$ state involves only one channel $^3P_0$. In contrast
with the $0^{++}$ and $1^{+-}$ cases, the tensor interaction
potential $-2V_{T}(r)$ is attractive as a first order effect instead
of a second order effect. The contributions of both spin-spin
interaction and spin-orbit interaction are attractive as well, since
$V_S(r)$ and $V_{LS}(r)$ are positive as shown in Fig.
\ref{potential}. The potential Eq.(\ref{6}) for this pseudoscalar is
displayed in Fig. \ref{single_channel_potential}a with
$\Lambda=1600$ MeV, we see that the potential is strong enough so
that the P wave centrifugal barrier can be partly compensated, then
there remains a weak attractive interaction in the intermediate
range. Therefore bound state solution can be found for reasonable
values of $\Lambda$, as can be seen from Table \ref{0-+}. For
$\Lambda=1500-1600$ MeV, we find the binding energy
$\varepsilon=1.40-114.81$ MeV. The binding energy is more sensitive
to $\Lambda$ than the $0^{++}$ and $1^{+-}$ two coupled channels
cases.

The results for $1^{++}$ state is similar to the single channel
$0^{-+}$ case. Because the D wave centrifugal barrier is higher than
the P wave centrifugal barrier, the total effective potential is
less attractive than the $0^{-+}$ state, this point can be seen
clearly in Fig. \ref{single_channel_potential}. If the coupling
constants are reduced by 20$\%$, bound state solution can be found
only for $\Lambda$ larger than 1920 MeV. $1^{++}$
$D^{*}_s\bar{D}^{*}_s$ is harder to be bound than the $0^{-+}$ state
due to the repulsive D wave centrifugal barrier.

We then come to the CP exotic $1^{-+}$ and $2^{+-}$ states, only one
channel is involved in both sectors. As has been shown in
Eq.(\ref{8}) and Eq.(\ref{9}), the potentials from one boson
exchange are exactly the same, and they are less attractive than the
potentials of the $0^{-+}$ and $1^{++}$ states. For the $1^{-+}$
configuration, bound state solution appears only for $\Lambda$ as
large as 2000 MeV, and we can find a $2^{+-}$ bound state only if
the regularization parameter $\Lambda$ is larger than 3160 MeV. If
we reduce all the coupling constants by 20$\%$, larger value of
$\Lambda$ is required to find bound state solutions. Because the
value of $\Lambda$ is so large that it is far beyond the range of
0.8 to 1.5 GeV favored by the nucleon-nucleon interactions, we tend
to conclude that the CP exotic $1^{-+}$ and $2^{+-}$
$D^{*}_s\bar{D}^{*}_s$ states can not be bound by the one boson
exchange potential. This conclusion is consistent with the fact that
no such CP exotic states have been observed so far.

For the $1^{--}$ states, there are three configurations $^1P_1$,
$^5P_1$ and $^5F_1$. In spite of the P wave centrifugal barrier,
bound state solutions can be found for reasonable value of
$\Lambda$, the reason is the large attractive contributions from the
tensor interaction and the spin-orbit interaction. From the
numerical results in Table \ref{1--}, we see that the binding energy
is rather sensitive to $\Lambda$, and $^5P_1$ is the dominant
component. This is because that the 22 element of the potential in
Eq. (\ref{10}) is more attractive than the 11 element, and the
$^5F_1$ component is strongly suppressed by the large F wave
centrifugal barrier. The $1^{--}$ $D^{*}_s\bar{D}^{*}_s$ state is
very interesting, it can be produced via the $e^{+}e^{-}$
annihilation or with the help of the initial state radiation (ISR)
technique at $B$ factory. The existence of such a state can be
confirmed or rejected, if more detailed $e^{+}e^{-}$ annihilation
data in the range of 4100-4200 MeV become available. We strongly
urge the Babar and Belle collaboration to search for this state, so
that our prediction can be checked. In addition to the dominant
decay modes $D^{+}_sD^{-}_s\gamma\gamma$ and
$D^{+}_sD^{-}_s\gamma\pi^{0}$, the $1^{--}$ $D^{*}_s\bar{D}^{*}_s$
molecule can also decay into $D_s\bar{D}_s$ and
$D_s\bar{D}^{*}_s/D^{*}_s\bar{D}_s$, which are other important decay
modes.

For the $2^{++}$ states, there are four channels $^1D_2$, $^5S_2$,
$^5D_2$ and $^5G_2$. We note that both the tensor interaction and
the spin-orbit interaction vanish in the $^5S_2$ configuration, and
the spin-spin interaction is repulsive. However, bound state
solution can be found for appropriate value of $\Lambda$ albeit its
value is somewhat larger than the corresponding ones in the
$0^{++}$, $1^{+-}$, $0^{-+}$ and $1^{--}$ cases. This is because
that the mixing of $^5S_2$ with $^1D_2$, $^5D_2$ and $^5G_2$ under
the tensor force increases the binding of the system through higher
order iterative processes. From the numerical results in Table
\ref{2++}, we see that the $^5S_2$ component is dominant, $^5D_2$
probability is larger than $^1D_2$, and $^5G_2$ has the smallest
probability.

Finally two states $2^{-+}$ and $2^{--}$ remain. For both states,
one has a two coupled channels (P wave and F wave configurations).
Because of the P wave and F wave centrifugal barrier, bound state
appears only for $\Lambda$ as large as 2080 MeV and 1890 MeV
respectively. Therefore both the $2^{-+}$ and $2^{--}$
$D^{*}_s\bar{D}^{*}_s$ states may not be bound by the one boson
exchange potential.

In short summary, ten allowed $D^{*}_s\bar{D}^{*}_s$ states with low
spin parity have been studied. We find that $J^{PC}=0^{++}$,
$1^{+-}$, $0^{-+}$, $2^{++}$ and $1^{--}$ states are more tightly
bound. We expect that the $0^{++}$ and $1^{+-}$
$D^{*}_s\bar{D}^{*}_s$ molecular states very likely exist, where the
system can be in S wave or D wave and the two configurations mix
with each other under the tensor interaction. Due to the remarkably
strong contributions from the spin-spin interaction, the tensor
interaction and the spin-orbit interaction, the $0^{-+}$
$D^{*}_s\bar{D}^{*}_s$ molecule may exist in spite of P wave
centrifugal barrier. The $1^{--}$ $D^{*}_s\bar{D}^{*}_s$ state
likely exist as well, and the $^5P_1$ component is dominant. One
expects that the one boson exchange potential could support the
$2^{++}$ $D^{*}_s\bar{D}^{*}_s$ bound state because of the presence
of the S wave configuration and mixing with the other three higher
partial waves. However, the CP exotic $1^{-+}$ and $2^{+-}$ states,
$2^{-+}$ and $2^{--}$ states should not exist.

\section{The interpretation of $Y(4140)$ as a $D^{*}_s\bar{D}^{*}_s$ molecule and its implication}

Since the one boson exchange potential does not support $1^{-+}$,
$2^{+-}$, $2^{-+}$ and $2^{--}$ $D^{*}_s\bar{D}^{*}_s$ molecular
states, we concentrate on $0^{++}$, $1^{+-}$, $0^{-+}$, $1^{++}$,
$1^{--}$ and $2^{++}$ states in the following. Because the $C-$
parity of $Y(4140)$ is positive, $J^{PC}=1^{+-}$ and $1^{--}$ states
are not possible. From the numerical results in Table
\ref{0++}-Table \ref{2--}, it is obvious that the $0^{++}$
$D^{*}_s\bar{D}^{*}_s$ is most tightly bound by the one boson
exchange potential. Consequently the most favorable quantum number
of $Y(4140)$ is $J^{PC}=0^{++}$, however, $J^{PC}=0^{-+}$ or
$2^{++}$ can not be excluded at present. It is crucial to perform a
partial wave analysis in future, if the spatial parity turns out to
be positive, $J^{PC}=0^{++}$ and $2^{++}$ are favored, otherwise it
may be a $0^{-+}$ state.

If $Y(4140)$ is confirmed to be a $0^{++}$ $D^{*}_s\bar{D}^{*}_s$
molecule with masses about 4143 MeV by future experiments, $0^{-+}$
and $2^{++}$ $D^{*}_s\bar{D}^{*}_s$ molecules should be observed as
well, whose masses are in the range of 4190 MeV to 4210 MeV. For the
parameters that allow $Y(4140)$ to emerge as a $0^{++}$
$D^{*}_s\bar{D}^{*}_s$ molecule, one expects that the $1^{+-}$ and
$1^{--}$ $D^{*}_s\bar{D}^{*}_s$ states may exist as well. We note
that the $C-$parity of both states is negative. The $1^{--}$ state
is particularly interesting, its mass is about 4120-4150 MeV, and it
can be produced directly in the $e^{+}e^{-}$ annihilation or via the
initial state radiation at $B$ factory. Both states can decay into
$D_s\bar{D}^{*}_s/D^{*}_s\bar{D}_s$, whereas the spin parity of the
first state forbids its decay into $D_s\bar{D}_s$. Since $Y(4140)$
is observed in the $J/\psi\phi$ channel, the $1^{+-}$ and $1^{--}$
states should be searched for in the $J/\psi\eta$ and $J/\psi\eta'$
final states. We suggest the CDF, Babar and Belle collaboration to
search for this state, which would be a critical check to the
$D^{*}_s\bar{D}^{*}_s$ molecule interpretation of $Y(4140)$.

If $Y(4140)$ is identified to be a $D^{*}_s\bar{D}^{*}_s$ molecule,
it is interesting to investigate whether a $D^{*}\bar{D}^{*}$
molecular state exists and what its most favorable quantum numbers
are. In Ref. \cite{Liu:2009ei}, the authors identified $Y(3940)$ as
the $D^{*}\bar{D}^{*}$ molecular partner of $Y(4140)$, then its
binding energy is about 77.5 MeV (Belle Collaboration) or 105.9 MeV
(Babar Collaboration). However, in one pion exchange model
T$\ddot{\rm o}$rnqvist demonstrated that the $D^{*}\bar{D}^{*}$
molecule should be near the threshold of 4020 MeV with
$J^{PC}=0^{++}$, $0^{-+}$, $1^{+-}$ or $2^{++}$
\cite{Tornqvist:1993ng}. It is necessary to reanalysis the
$D^{*}\bar{D}^{*}$ system dynamically. In Ref.\cite{Ding:2009zq},
the contributions from $\pi$, $\eta$, $\sigma$, $\rho$ and $\omega$
exchanges are included, the binding energy and other static
properties are found to be sensitive to the regularization parameter
$\Lambda$. However, we qualitatively confirmed that the
$J^{PC}=2^{++}$, $1^{--}$, $0^{++}$ and $0^{-+}$ $D^{*}\bar{D}^{*}$
states are more deeply bound. Consequently its huge binding energy
implies the existence of more $D^{*}\bar{D}^{*}$ molecules with
different quantum numbers, if $Y(3940)$ is identified as a
$D^{*}\bar{D}^{*}$ bound state. However, no such candidates have
been reported experimentally so far, therefore we tend to conclude
that $Y(3940)$ as a $D^{*}\bar{D}^{*}$ molecule is not favored.
Further experimental and theoretical efforts are needed to
understand the structure and the properties of $Y(3940)$. With the
same argument as that in the introduction section, $Y(3940)$ as a
canonical $c\bar{c}(2P)$ charmonium can not be completely excluded.
The charmonium assignment of $Y(3940)$ can be tested by searching
for $D\bar{D}$ and $D\bar{D}^{*}/D^{*}\bar{D}$ final states and by
studying their angular distributions \cite{Godfrey:2008nc}.

For the $B^{*}_s\bar{B}^{*}_s$ system, the repulsive kinetic energy
is greatly reduced due to larger mass of $B^{*}_s$ meson, therefore
the $B^{*}_s\bar{B}^{*}_s$ state should be more deeply bound than
$D^{*}_s\bar{D}^{*}_s$. We expect that at least $0^{++}$
$B^{*}_s\bar{B}^{*}_s$ molecular state should exist with larger
binding energy. Obviously this state could be searched for in the
$\Upsilon(1S)\phi$ channel. Because of its large mass, the most
promising places to produce this state conspicuously are the large
hadron colliders such as Tevatron and LHC.

\section{conclusions and discussions}

In this work, we have dynamically studied the possible
$D^{*}_s\bar{D}^{*}_s$ molecular states and the interpretation of
$Y(4140)$ as a $D^{*}_s\bar{D}^{*}_s$ molecule in the one boson
exchange model, where $\sigma$, $\eta$ and $\phi$ exchanges are
taken into account. Ten allowed states with low spin parity have
been considered, we would like to stress that only S wave
configuration is usually considered in the familiar phenomenological
models such as the one boson exchange model in heavy quark effective
theory \cite{zhu_liu,Ding:2008gr} and the potential model with
pairwise interactions \cite{Wong:2003xk,Ding:2008mp}. We find that
the binding energy and static properties are sensitive to the
regularization parameter $\Lambda$ and the effective coupling
constants. Since the regularization parameter $\Lambda$ is poorly
known so far, we are not able to precisely predict the binding
energies of the possible $D^{*}_s\bar{D}^{*}_s$ molecular states
bound by one boson exchange potential. However, we can reliably
predict which ones of the ten allowed states are much easier to be
bound, and the predictions are rather stable even if the uncertainty
of the coupling constants is considered, as is obvious from the
numerical results listed in the manuscript. Further research on
X(3872), which is a promising ${D\bar{D}^{*}/\bar{D}D^{*}}$
molecule, would severely constraint the parameters of the one boson
exchange model, especially the regularization parameter $\Lambda$,
so that the predictions presented in the work could become more
precise.

We quantitatively confirm that the $0^{++}$ $D^{*}_s\bar{D}^{*}_s$
state is most easily to be bound. Our numerical results imply that
the $J^{PC}=0^{++}$, $1^{+-}$, $0^{-+}$, $2^{++}$ and $1^{--}$
configurations are rather more strongly bound so that the
corresponding $D^{*}_s\bar{D}^{*}_s$ molecules may exist, whereas
the CP exotic $1^{-+}$ and $2^{+-}$, $2^{-+}$ and $2^{--}$
$D^{*}_s\bar{D}^{*}_s$ states are not be bound by the one boson
exchange potential. We note that the possible existence of a number
of bound state is not a specific prediction of our model
\cite{zhu_liu,Tornqvist:1993ng}. Finally we would like to stress
that we still can not completely rule out $Y(4140)$ as a
conventional $c\bar{c}$ charmonium at present, in spite of its
peculiar decay mode $J/\psi\phi$. From the theoretical predictions
for the charmonium spectrum, $Y(4140)$ is most likely to be the
$2\,^1D_2$ state with $I^{G}(J^{PC})=0^{+}(2^{-+})$, if it is a
$c\bar{c}$ charmonium state. It unusual large branch ratio into
$J/\psi\phi$ may be explained by the rescattering mechanism or the
mixing between charmonium and molecule. Compared with other
charmonium like states, the experimental information for $Y(4140)$
is scarce, and further experiment data are critically needed.

For $Y(4140)$ as a $D^{*}_s\bar{D}^{*}_s$ molecule, we suggest that
its most favorable quantum numbers are $J^{PC}=0^{++}$, although
$J^{PC}=0^{-+}$ and $2^{++}$ can not be ruled out by the present
experimental data. It mainly decays into
$D^{+}_sD^{-}_s\gamma\gamma$ and $D^{+}_sD^{-}_s\gamma\pi^{0}$ via
almost free decay of $D^{*}_s$ and $\bar{D}^{*}_s$, and the decay
mode $D^{+}_sD^{-}_s\pi^{0}\pi^{0}$ is forbidden by phase space
constraints. The search for the four body decays $Y(4140)\rightarrow
D^{+}_sD^{-}_s\gamma\gamma$ and $Y(4140)\rightarrow
D^{+}_sD^{-}_s\gamma\pi^{0}$ is crucial to test the hadronic
molecule hypothesis of $Y(4140)$. If $Y(4140)$ is confirmed to be a
$0^{++}$ $D^{*}_s\bar{D}^{*}_s$ molecule by future theoretical and
experimental efforts, the $0^{-+}$ and $2^{++}$ partners should
exist with mass in the range of 4190-4210 MeV. We argue that the
$1^{+-}$ and $1^{--}$ $D^{*}_s\bar{D}^{*}_s$ states with negative
$C-$parity should be observed as well. The $1^{--}$ states can be
produced largely in the $e^{+}e^{-}$ annihilation or with the help
the initial state radiation at $B$ factory, and detailed
$e^{+}e^{-}$ annihilation data near 4100$\sim$4200 MeV are important
to confirm or refute the existence of such state. Both the $1^{+-}$
and $1^{--}$ $D^{*}_s\bar{D}^{*}_s$ states can be searched for in
the $J/\psi\eta$ and $J/\psi\eta'$ final states. We strongly urge
the CDF, Babar and Belle Collaborations to search for these two
negative $C-$parity states, which would be another important test to
the molecular hypothesis of $Y(4140)$ and the reliability of our one
boson exchange model.

If we identify $Y(3940)$ as a $D^{*}\bar{D}^{*}$ molecule, its large
binding energy requires the existence of more $D^{*}\bar{D}^{*}$
bound states, which has not been observed so far. Therefore the
interpretation of $Y(3940)$ as a $D^{*}\bar{D}^{*}$ molecule may not
be favored in our opinion. With the present experiment data, the
charmonium assignment for $Y(3940)$ can not be ruled out. We suggest
that the $0^{++}$ $B^{*}\bar{B}^{*}_s$ molecular state should exist,
and it is bound more tightly than $D^{*}_s\bar{D}^{*}_s$. We should
search for this state at Tevatron or LHC in the $\Upsilon(1S)\phi$
channel.

\begin{acknowledgments}
We are grateful to Prof. Mu-Lin Yan and Dao-Neng Gao for stimulating
discussions. This work is supported by the China Postdoctoral
Science foundation (20070420735), K.C. Wong Education Foundation,
and KJCX2-YW-N29 of the Chinese Academy.
\end{acknowledgments}


\begin{appendix}
\section{Numerical results for the ten allowed $D^{*}_s\bar{D}^{*}_s$ states\label{appendix}}


\begin{center}
\begin{table}[hptb]
\begin{tabular}{|c|ccc|}\hline\hline

$\Lambda({\rm MeV})$&$~~~{\varepsilon}(\rm MeV)$&$~~~{\rm r}_{\rm
rms}({\rm fm})$&$~~~{\rm P_S:P_D(\%)}$\\\hline

1350&   6.46  & 1.49   &99.48:0.52   \\
1400&  16.09  & 1.03  & 98.83:1.17  \\
1450&  32.11  & 0.79   & 97.51:2.49  \\
1500&   57.33 & 0.63   &  94.56:5.44     \\
1550&  97.73  &  0.52  & 87.27:12.73    \\
1600&  168.73   & 0.44   & 69.44:30.56
\\\hline\hline

\multicolumn{4}{|c|}{all couplings are reduced by 20 percents
}\\\hline\hline

$\Lambda({\rm MeV})$&$~~~{\varepsilon}(\rm MeV)$&$~~~{\rm r}_{\rm
rms}({\rm fm})$&$~~~{\rm P_S:P_D}(\%)$\\
\hline

1500 & 4.20  &   1.75& 99.51:0.49 \\
1550 &  10.28 & 1.20 & 98.94:1.06 \\
1600 & 20.14  & 0.91  & 97.88:2.13 \\
1650 & 35.21  & 0.73  &95.77:4.22  \\
1700 & 58.25  & 0.60  &91.35:8.65  \\
1750 & 95.26 &  0.50 & 81.68:18.32 \\ \hline \hline

\end{tabular}
\caption{\label{0++}The predictions for the static properties of the
${\rm J^{PC}=0^{++}}$ ${\rm D^{*}_s\bar{D}^{*}_s}$ hadronic
molecule, where $\varepsilon$ denotes the binding energy, rms is the
root of mean square radius, ${\rm P_S}$ and ${\rm P_D}$ represent
the S state and D state probabilities respectively.}
\end{table}
\end{center}


\begin{center}
\begin{table}[hptb]
\begin{tabular}{|c|ccc|}\hline\hline

$\Lambda({\rm MeV})$&$~~~{\rm \varepsilon}(\rm MeV)$&$~~~{\rm
r}_{\rm rms}({\rm fm})$&$~~~{\rm P_S:P_D(\%)}$\\\hline

1350&  6.13 &  1.53 & 99.65:0.35 \\
1400& 13.34  &  1.12  & 99.36:0.64 \\
1450& 23.76   &  0.89 & 98.95:1.05 \\
1500& 37.74   & 0.75  & 98.35:1.65 \\
1550& 55.74  & 0.65  & 97.48:2.52 \\
1600& 78.42  & 0.58  & 96.16:3.84 \\
1650&  106.87 & 0.52  & 94.15:5.85 \\\hline\hline

\multicolumn{4}{|c|}{all couplings are reduced by 20
percents}\\\hline\hline

$\Lambda({\rm MeV})$&$~~~{\varepsilon}(\rm MeV)$&$~~~{\rm r}_{\rm
rms}({\rm fm})$&$~~~{\rm P_S:P_D}(\%)$\\
\hline

1600 & 10.55   & 1.20  & 99.29:0.71 \\
1650 & 17.09   & 0.98 & 98.91:1.09 \\
1700 &25.48   & 0.84  & 98.39:1.61 \\
1750 & 35.97   &0.74   & 97.67:2.33 \\
1800 & 48.92   & 0.66  & 96.67:3.33 \\
1850 & 64.84  & 0.59  &95.27:4.73  \\
1900 & 84.49  & 0.54  &93.29:6.71  \\
1950 & 109.01  & 0.49  & 90.48:9.52 \\
\hline \hline

\end{tabular}
\caption{\label{1+-}The predictions about the binding energy, the
root of mean square radius(rms) and the probabilities of the
different components for the $1^{+-}$ ${\rm D^{*}_s\bar{D}^{*}_s}$
molecule.}
\end{table}
\end{center}


\begin{center}
\begin{table}[hptb]
\begin{tabular}{|c|cc|}\hline\hline

$\Lambda({\rm MeV})$&$~~~{\rm \varepsilon}(\rm MeV)$&$~~~{\rm
r}_{\rm rms}({\rm fm})$\\\hline

1500& 1.40   &   1.52  \\
1520& 14.99   &  0.86   \\
1540& 33.36   &  0.70   \\
1560&  56.13  &  0.62   \\
1580& 83.26   &   0.56  \\
1600& 114.81   &   0.51  \\\hline\hline

\multicolumn{3}{|c|}{all couplings are reduced by 20
percents}\\\hline\hline

$\Lambda({\rm MeV})$&$~~~{\rm M}(\rm MeV)$&$~~~{\rm r}_{\rm
rms}({\rm fm})$\\
\hline

1700 & 6.56  &  0.99  \\
1720 & 19.69  &  0.76  \\
1740 & 35.86  &  0.65  \\
1760 & 54.89  &  0.58  \\
1780 & 76.74  &  0.54  \\
1800 &  101.41  & 0.50  \\\hline \hline

\end{tabular}
\caption{\label{0-+}The predictions for the binding energy and the
rms of the $0^{-+}$ ${\rm D^{*}_s\bar{D}^{*}_s}$ molecule.}
\end{table}
\end{center}


\begin{center}
\begin{table}[hptb]
\begin{tabular}{|c|cc|}\hline\hline

$\Lambda({\rm MeV})$&$~~~{\varepsilon}(\rm MeV)$&$~~~{\rm r}_{\rm
rms}({\rm fm})$\\\hline

1680& 4.21   &   0.62  \\
1690& 28.16   &   0.54   \\
1700& 54.83   &   0.50  \\
1710& 84.04   &   0.47  \\
1720& 115.71    &   0.45   \\\hline\hline

\multicolumn{3}{|c|}{all couplings are reduced by 20
percents}\\\hline\hline

$\Lambda({\rm MeV})$&$~~~{\varepsilon}(\rm MeV)$&$~~~{\rm r}_{\rm
rms}({\rm fm})$\\
\hline

1920 & 12.37  &   0.54  \\
1930 & 33.19  &   0.49  \\
1940 & 55.70  &   0.46  \\
1950 &79.81  &  0.44  \\
1960 & 105.50  &  0.43   \\\hline \hline

\end{tabular}
\caption{\label{1++}The predictions for the binding energy and the
rms of the $1^{++}$ ${\rm D^{*}_s\bar{D}^{*}_s}$ molecule.}
\end{table}
\end{center}


\begin{center}
\begin{table}[hptb]
\begin{tabular}{|c|cc|}\hline\hline

$\Lambda({\rm MeV})$&$~~~{\varepsilon}(\rm MeV)$&$~~~{\rm r}_{\rm
rms}({\rm fm})$\\\hline

2000& 10.72  &  0.77  \\
2020& 23.60  & 0.63   \\
2040& 38.60  &  0.56 \\
2060& 55.61  & 0.51   \\
2080& 74.60  & 0.47  \\
2100& 95.54  &  0.44  \\
2120& 118.44  & 0.42   \\\hline\hline

\multicolumn{3}{|c|}{all couplings are reduced by 20
percents}\\\hline\hline

$\Lambda({\rm MeV})$&$~~~{\varepsilon}(\rm MeV)$&$~~~{\rm r}_{\rm
rms}({\rm fm})$\\
\hline

2480 & 9.94   & 0.72 \\
2500 & 20.67 &   0.60\\
2520 & 32.81  &  0.53  \\
2540 &  46.25 &  0.49  \\
2560 &  60.96 &  0.45  \\
2580 & 76.92  &  0.42  \\
2600 & 94.11 &  0.40 \\
2620 & 112.54  & 0.38   \\\hline \hline

\end{tabular}
\caption{\label{1-+}The predictions for the binding energy and the
rms of the $1^{-+}$ ${\rm D^{*}_s\bar{D}^{*}_s}$ molecule.}
\end{table}
\end{center}


\begin{center}
\begin{table}[hptb]
\begin{tabular}{|c|cc|}\hline\hline

$\Lambda({\rm MeV})$&$~~~{\varepsilon}(\rm MeV)$&$~~~{\rm r}_{\rm
rms}({\rm fm})$\\\hline

3160 & 6.48   &   0.37  \\
3170 & 22.41   &   0.34  \\
3180 & 39.00   &   0.33 \\
3190 & 56.20   &  0.32   \\
3200 & 74.00   &   0.31  \\
3210 & 92.40   &   0.30  \\
3220 & 111.38  &   0.29  \\\hline\hline

\multicolumn{3}{|c|}{all couplings are reduced by half
}\\\hline\hline

$\Lambda({\rm MeV})$&$~~~{\varepsilon (\rm MeV)}$&$~~~{\rm r}_{\rm
rms}({\rm fm})$\\
\hline

4420 & 11.51  & 0.28   \\
4430 & 28.97  & 0.26   \\
4440 & 46.92  & 0.25   \\
4450 & 65.36  & 0.25  \\
4460 & 84.27  & 0.24  \\
4470 & 103.63 & 0.23   \\\hline \hline

\end{tabular}
\caption{\label{2+-}The predictions for the binding energy and the
rms of the $2^{+-}$ ${\rm D^{*}_s\bar{D}^{*}_s}$ molecule.}
\end{table}
\end{center}


\begin{center}
\begin{table}[hptb]
\begin{tabular}{|c|ccc|}\hline\hline

$\Lambda({\rm MeV})$&$~~~{\varepsilon}(\rm MeV)$&$~~~{\rm r}_{\rm
rms}({\rm fm})$&$~~~{\rm P_{P0}:P_{P2}:P_{F}(\%)}$\\\hline

1470 & 5.38   & 1.10 &3.40:96.50:0.10   \\
1480 & 13.13   & 0.89 & 3.18:96.71:0.11  \\
1490 & 22.23   & 0.78 & 3.01:96.87:0.13   \\
1500 & 32.59   &  0.71 & 2.86:97.00:0.14  \\
1510 & 44.18   & 0.66 & 2.73:97.12:0.16  \\
1520 & 56.97    &  0.62  & 2.61:97.22:0.17  \\
1530 & 70.95   & 0.59 & 2.51:97.30:0.19  \\
1540 & 86.12   & 0.56 & 2.42:97.38:0.20   \\
1550 &  102.49  &  0.53 & 2.34:97.44:0.22
\\\hline\hline

\multicolumn{4}{|c|}{all couplings are reduced by 20
percents}\\\hline\hline

$\Lambda({\rm MeV})$&$~~~{\varepsilon}(\rm MeV)$&$~~~{\rm r}_{\rm
rms}({\rm fm})$&$~~~{\rm P_{P0}:P_{P2}:P_{F}}(\%)$\\
\hline

1650 & 11.28  &  0.88 &2.46:97.45:0.09 \\
1660 &18.49   &0.78   & 2.36:97.53:0.11 \\
1670 & 26.53  & 0.71  & 2.29:97.60:0.12 \\
1680 & 35.36  & 0.67  & 2.22:97.66:0.13 \\
1690 & 44.96  & 0.63  &2.15:97.71:0.14  \\
1700 &  55.33 & 0.60  & 2.10:97.75:0.15\\
1710 & 66.45  & 0.57  & 2.04:97.79:0.16 \\
1720 & 78.32  & 0.55  & 2.00:97.83:0.18 \\
1730 &90.95   &  0.53 &1.95:97.86:0.19  \\
1740 &104.33 & 0.51  & 1.91:97.89:0.20
\\\hline \hline

\end{tabular}
\caption{\label{1--}The predictions about the binding energy, the
root of mean square radius and the probabilities of the different
components for the $1^{--}$ ${\rm D^{*}_s\bar{D}^{*}_s}$ molecule,
where ${\rm P_{P0}}$ and ${\rm P_{P2}}$ denote the ${\rm ^1P_1}$
state and ${\rm ^5P_1}$  state probabilities respectively.}
\end{table}
\end{center}


\begin{center}
\begin{table}[hptb]
\begin{tabular}{|c|ccc|}\hline\hline

$\Lambda({\rm MeV})$&$~~~{\varepsilon}(\rm MeV)$&$~~~{\rm r}_{\rm
rms}({\rm fm})$&$~~~{\rm P_{D0}:P_{S}:P_{D2}:P_{G}(\%)}$\\\hline

1350& 6.81   & 1.47   &   0.07:99.48:0.46:0.00 \\
1400&  12.03  &  1.18  & 0.10:99.14:0.75:0.00    \\
1450&  18.59  &  1.00 &  0.15:98.67:1.18:0.00   \\
1500&  26.55  & 0.88    & 0.21:98.01:1.78:0.00   \\
1550&  36.10   & 0.80   & 0.29:97.07:2.63:0.00    \\
1600&  47.59  &  0.73  &  0.40:95.74:3.85:0.01   \\
1650&  61.65  & 0.67   &  0.53:93.85:5.60:0.02   \\
1700& 79.28  &  0.62   &  0.70:91.12:8.13:0.05   \\
1750&  102.10 &  0.57 & 0.91:87.18:11.77:0.14
\\\hline\hline

\multicolumn{4}{|c|}{all couplings are reduced by 20 percents
}\\\hline\hline

$\Lambda({\rm MeV})$&$~~~{\varepsilon}(\rm MeV)$&$~~~{\rm r}_{\rm
rms}({\rm fm})$&$~~~{\rm P_{D0}:P_{S}:P_{D2}:P_{G}}(\%)$\\
\hline

1650  & 8.09 &  1.36&  0.11:98.92:0.97:0.00 \\
1700  & 11.69  & 1.18  & 0.15:98.44:1.40:0.00\\
1750  & 16.09  & 1.05  & 0.21:97.81:1.98:0.00\\
1800  &  21.41 & 0.94  &0.27:96.97:2.75:0.01 \\
1850  & 27.91  & 0.86  &0.35:95.86:3.79:0.01 \\
1900  & 35.91  & 0.78  &0.44:94.36:5.17:0.02  \\
1950  & 45.91  &  0.72 &  0.56:92.36:7.04:0.04 \\
2000  &  58.62 & 0.66& 0.70:89.64:9.56:0.09  \\
2050  &  75.18 & 0.60  & 0.89:85.95:12.95:0.22\\
2100  & 97.42  & 0.55  & 1.13:80.82:17.48:0.57\\
2150 & 129.26 & 0.50  & 1.49:73.08:23.51:1.93
\\\hline \hline

\end{tabular}
\caption{\label{2++}The predictions about the binding energy, the
root of mean square radius and the probabilities of the different
components for the $2^{++}$ ${\rm D^{*}_s\bar{D}^{*}_s}$ molecule.}
\end{table}
\end{center}


\begin{center}
\begin{table}[hptb]
\begin{tabular}{|c|ccc|}\hline\hline

$\Lambda({\rm MeV})$&$~~~{\varepsilon}(\rm MeV)$&$~~~{\rm r}_{\rm
rms}({\rm fm})$&$~~~{\rm P_P:P_F(\%)}$\\\hline

2080&  21.35  & 0.54   & 32.02:67.98  \\
2090&  50.76  & 0.46   & 25.30:74.70  \\
2100& 84.10   & 0.42   &   21.08:78.92 \\
2110& 120.80   & 0.39   & 18.09:81.91
\\\hline\hline

\multicolumn{4}{|c|}{all couplings are reduced by 20 percents
}\\\hline\hline

$\Lambda({\rm MeV})$&$~~~{\varepsilon}(\rm MeV)$&$~~~{\rm r}_{\rm
rms}({\rm fm})$&$~~~{\rm P_P:P_F}(\%)$\\
\hline

2510 & 18.99  &  0.47  &24.99:75.01  \\
2520 &  47.32&  0.40 & 20.26:79.74 \\
2530 & 78.24  & 0.37  &17.38:82.62 \\
2540 & 111.36  & 0.35  & 15.34:84.66 \\\hline \hline

\end{tabular}
\caption{\label{2-+}The predictions about the binding energy, the
root of mean square radius and the probabilities of the different
components for the $2^{-+}$ ${\rm D^{*}_s\bar{D}^{*}_s}$ molecule.}
\end{table}
\end{center}


\begin{center}
\begin{table}[hptb]
\begin{tabular}{|c|ccc|}\hline\hline

$\Lambda({\rm MeV})$&$~~~{\varepsilon}(\rm MeV)$&$~~~{\rm r}_{\rm
rms}({\rm fm})$&$~~~{\rm P_P:P_F(\%)}$\\\hline

1890&  1.59   &  1.12  & 65.79:34.21   \\
1900&  21.17   & 0.61   & 51.31:48.69  \\
1910&  46.97   & 0.51   & 41.55:58.46  \\
1920& 78.03    & 0.46    & 34.08:65.92 \\
1930& 113.83    & 0.42   &     28.31:71.69  \\\hline\hline

\multicolumn{4}{|c|}{all couplings are reduced by 20
percents}\\\hline\hline

$\Lambda({\rm MeV})$&$~~~{\varepsilon}(\rm MeV)$&$~~~{\rm r}_{\rm
rms}({\rm fm})$&$~~~{\rm P_P:P_F}(\%)$\\
\hline

2230 &  5.92 & 0.71  &49.37:50.63  \\
2240 & 28.13  & 0.50  & 38.73:61.27 \\
2250 & 54.53  &  0.44 & 31.95:68.05 \\
2260 & 84.34  & 0.40 & 26.99:73.01 \\
2270 &  117.14  & 0.38 & 23.20:76.80 \\\hline \hline

\end{tabular}
\caption{\label{2--}The predictions about the binding energy, the
root of mean square radius and the probabilities of the different
components for the $2^{--}$ ${\rm D^{*}_s\bar{D}^{*}_s}$ molecule.}
\end{table}
\end{center}

\end{appendix}

\end{document}